\documentclass{article}
\usepackage{graphicx}

\begin{document}

\title{Status of the TREND project}
\author{Olivier Martineau-Huynh for the TREND collaboration\\
{\small  Laboratoire de Physique Nucl\'eaire et des Hautes Energies,} \\
{\small  CNRS/IN2P3 and Universit\'e Pierre et Marie Curie, Paris, France,} \\
{\small Institute of High Energy Physics, Chinese Academy of  Sciences, Beijing, China, and
} \\
{\small National Astronomical Observatories of China, Chinese Academy of  Sciences, Beijing, China} \\
{\small {\it Email} : omartino@in2p3.fr}}


\date{}
\maketitle
\begin{abstract}
The Tianshan Radio Experiment for Neutrino Detection (TREND) is a sino-french collaboration (CNRS/IN2P3 and Chinese Academy of Science) developing an autonomous antenna array for the detection of high energy Extensive Air Showers (EAS) on the site of the 21CMA radio observatory. The autonomous detection and identification of EAS was achieved by TREND on a prototype array in 2009. This result was confirmed soon after when EAS radio-candidates could be tagged as cosmic ray events by an array of particle detectors running in parallel at the same location. This result is an important milestone for TREND, and more generally, for the maturation of the EAS radio-detection technique. The array is presently composed of 50 antennas covering a total area of $\approx1.2$~km$^2$, running in steady conditions since March 2011. We are presently processing these data to identify EAS radio-candidates. In a long term perspective, TREND is intended to search for high energy tau neutrinos. Here we only report on the results achieved so far by TREND.
\end{abstract}

\section{Introduction}
 Charged particles created during the development of an extensive air shower (EAS) generate a coherent emission detectable at radio frequencies ($<$ 200~ MHz). The radio-detection technique shows interesting potentialities for the detection of cosmic rays with energies above 10$^{17}$~eV (low cost, easiness of deployment and stability of the response of the antennas,...) and pioneering experiments \cite{cod1},\cite{lop1} obtained very encouraging results on the radio-detection of EAS. Still, the radio technique is not yet mature: more data is necessary to better understand the mechanism of shower generation, and self-triggering of the radio array is a necessary technical improvement. Among several other projects \cite{revenu}, TREND aims at tackling these two issues. We briefly report on this effort here.

\section{The TREND setup}
\begin{figure}
\includegraphics[width=6cm,height=4.5cm]{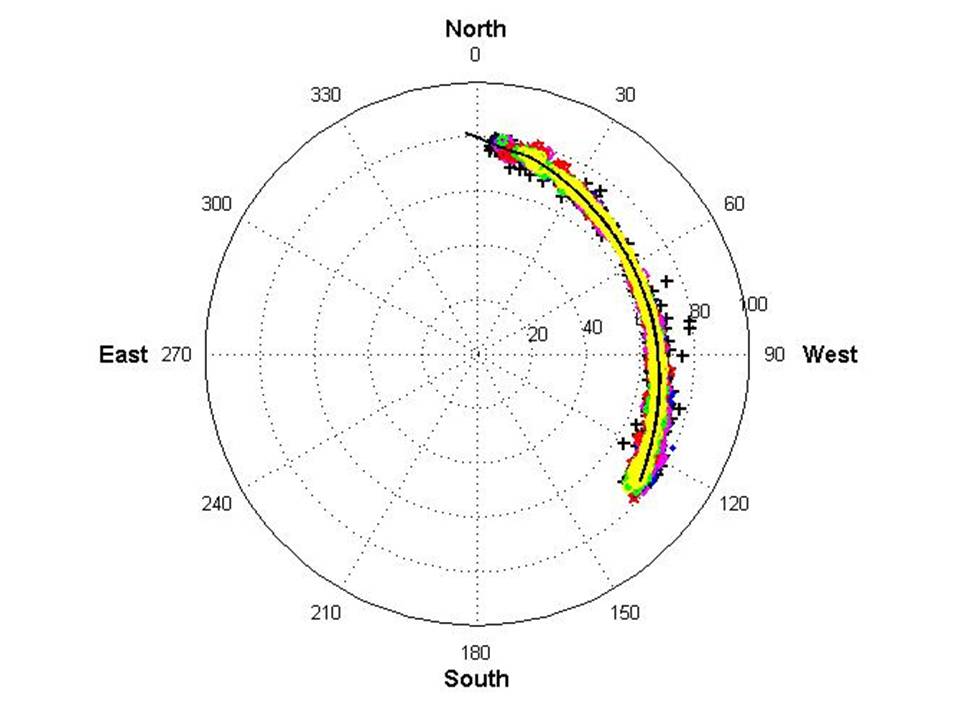}
\includegraphics[width=6cm,height=4.5cm]{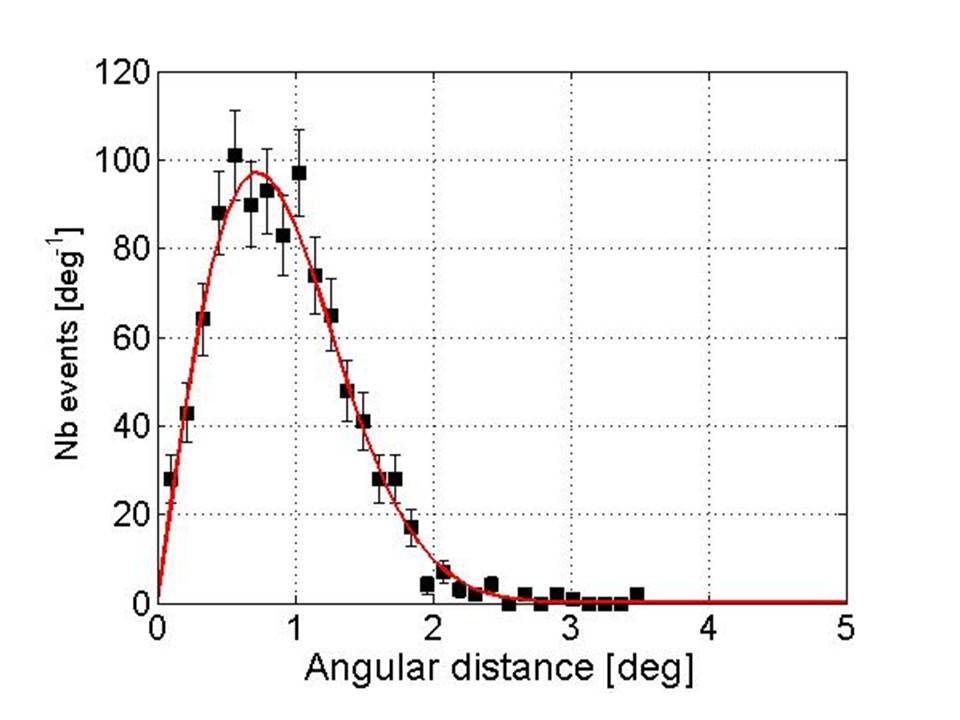}

\caption{
\label{fig_1}
\textit{
Left: skyplot of a plane trajectory reconstructed by TREND. Right: histogram of the angular distance between the estimated and reconstructed plane positions for the events shown in the left panel.
}
}
\end{figure}

The TREND setup was developed on the site of the 21cm Array radio telescope~\cite{21cma}, in the Tianshan mountain range (Xinjiang autonomous province, China). It benefits from the infrastructure of the 21CMA and its exceptional radio environment. A 6-antennas array could be promptly deployed in 2009 by adapting elements of the 21CMA setup. This prototype was used to test the principle of EAS autonomous radio-detection. The array was then extended to 15 antennas, while a setup of 3 particle detectors was also installed at the same location. An offline analysis of 29 live days of hybrid data showed that 13 EAS candidates selected from the radio data were coincident with cosmic ray events identified with the particle detectors setup. This validates our principle of radio-identification of EAS. These results are detailed in \cite{astro}.\\
The TREND array was then extended to 50 antennas, for a total area of $\approx$1.2~km$^2$. The system has been running in stable condition since March 2011. The good performance of the array could be checked through the reconstruction of plane trajectories in the sky (see figure \ref{fig_1}), with an angular resolution of $\approx 1^o$ and an amplitude resolution of 15\%.

\section{EAS radio-identification with TREND}
\begin{figure}[h]
\begin{center}
\includegraphics[width=10cm]{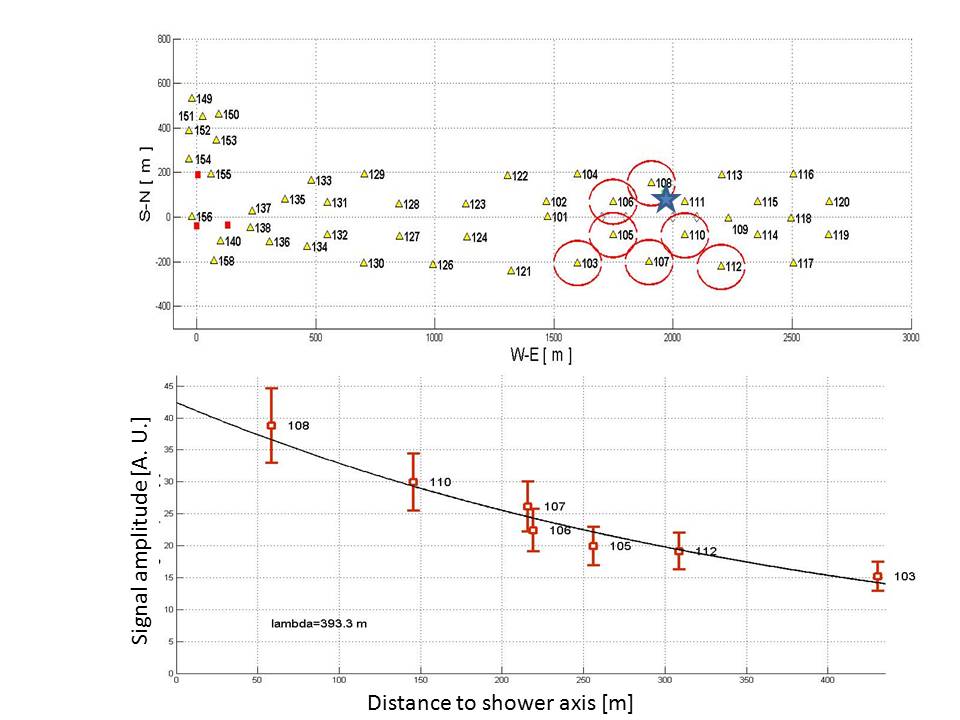}
\caption{Top: TREND 50-antennas array. Triangles are antennas, squares are particle detectors. Also shown with circles is the pattern at ground of an EAS radio-candidate. The incoming wave is reconstructed as a plane with $\theta=50.5^o$ and $\phi=4.2^o$. The reconstructed core position is shown as a blue star. Bottom: antennas signal amplitudes as a function of the distance to the reconstructed shower axis, and exponential fit ($\lambda$ = 393~m).}
\label{fig_2}
\end{center}
\end{figure}

EAS radio candidates are supposed to exhibit features distinct from background electromagnetic events. Prompt ($<$300~ns) electromagnetic transients are expected, as well as plane wavefronts. Well-focused patterns on ground, with a rapid drop in amplitude when moving away from the shower axis, is another characteristics of EAS radio signals.
The search for such signatures is implemented in the algorithm we developed for the selection of EAS radio-candidates. This algorithm is
presently being refined to be applied to the full 50-antennas array data, but EAS radio-candidates could already be identified (figure \ref{fig_2}).

\section{Conclusion}
The TREND project has been able to identify EAS candidates through their radio signatures only. We are confident that the 50-antennas array will allow to collect large statistics of EAS radio-events. This should prove helpful to better understand and control this promising technique.

\end{document}